\documentclass[twocolumn,superscriptaddress,showpacs,preprintnumbers,amsmath,amssymb,pre]{revtex4}
\usepackage{graphicx}
\usepackage{dcolumn,xcolor,ulem}
\usepackage{amsmath} 
\usepackage{amssymb}
\usepackage{amsfonts}
\usepackage{bm}
\usepackage[latin1]{inputenc}

\newcommand\gp{\dot\gamma}

\begin{document}

%\widowpenalty=1000
%\clubpenalty=1000
%\preprint{APS/123-QED}

 \title{Shear Banding of Complex Fluids}

\author{Thibaut Divoux}
\affiliation{Centre de Recherche Paul Pascal, UPR 8641, 115 av. Dr. Schweitzer, 33600 Pessac,
France}
\author{Marc A. Fardin}
\affiliation{Institut Jacques Monod, CNRS UMR 7592, Universit\'e Paris-Diderot, 15 rue H\'el\`ene Brion, 75205 Paris C\'edex 13, France}
\author{S\'ebastien Manneville}
\affiliation{Universit\'e de Lyon, Laboratoire de Physique, \'Ecole Normale Sup\'erieure de Lyon, CNRS UMR 5672, 46 All\'ee d'Italie, 69364 Lyon cedex 07, France}
\author{Sandra Lerouge}
\affiliation{Laboratoire Mati\`ere et Syst\`emes Complexes, CNRS UMR 7057, Universit\'e Paris-Diderot, 10 rue Alice Domon et L\'eonie Duquet, 75205 Paris C\'edex 13, France}

\date{\today}

\begin{abstract}
Even in simple geometries many complex fluids display non-trivial flow fields, with regions where shear is concentrated. The possibility for such shear banding has been known since several decades, but the recent years have seen an upsurge of studies offering an ever more precise understanding of the phenomenon. The development of new techniques to probe the flow on multiple scales and with increasing spatial and temporal resolution has opened the possibility for a synthesis of the many phenomena that could only have been thought of separately before. In this review, we  bring together recent research on shear banding in polymeric and on soft glassy materials, and highlight their similarities and disparities.
\end{abstract}

\maketitle

\section{Introduction}

Shear banding seems to have been identified first at the end of the 19th century by geologists and engineers working on the deformation of solids~\cite{Wright:2002}. In this context, shear bands are zones where the strain $\gamma$ can take values much larger than in the rest of the sample. The present review focuses on shear banding in complex fluids which are part of what one may call ``soft matter''~\cite{PGG:1992}: a somewhat overlapping collection of systems such as liquid crystals, colloidal dispersions, polymer solutions, melts or gels, emulsions, pastes, foams, etc.~\cite{Larson:1999}. In all these systems, the constituents are often said to be ``mesoscopic'' since the relevant length scale, which in practice is not always clearly identified, ranges between the molecular scale and the flow scale. Macroscopically, the mechanical properties of soft matter sit in between those of the ideal Hookean elastic solid, and those of the ideal Newtonian viscous fluid. In this context, the notion of shear banding is associated with the localization of the shear (or strain) rate $\gp$.

Shear banding is ubiquitous in complex fluids, and has been reported in systems such as wormlike micellar surfactant solutions, lyotropic lamellar surfactant phases, polymer solutions and melts, star polymers, emulsions, suspensions, microgels, biological gels and foams. Several reviews have been published in recent years concerning various theoretical and experimental aspects. For instance, in the \textit{Annual Reviews of Fluid Mechanics}, the topic was last approached by \cite{Goddard:2003} --from a mechanistic perspective grounded in the historical background on shear banding in solids, and more recently by \cite{Schall:2010}-- whom mainly focused on recent experiments on shear banding in granular matter. Over the five years since, more than five hundred articles mentioning ``shear banding'' and ``complex fluids'' have been published. This profusion of studies reflects the growing interest for this subject, and justifies the need for a new review today. This profusion is also a consequence of the very broad --but sometimes loose-- use of the term ``shear banding''. The words chosen in particular research trends may be somewhat contextual and determining whether two words across fields are synonyms or not is not always easy. Some of what other authors have called ``layers'' or ``stripes'' or even ``kinks'', and ``fractures'' may be close to our use of the term ``bands'' in this review; and vice-versa. Our goal is not to authoritatively define what shear banding is and is not, but to witness the evolution of, and refinement over the use of the term ``shear banding'' in recent soft matter research. 

In practice, we shall not cover the five hundred references of the last five years but we will focus on two classes of materials: polymeric fluids that will be discussed in section~\ref{polym}, and soft glassy materials that will be discussed in section~\ref{softglass}. The first class of materials will sit more on the side of fluids, whereas the second will sit more on the side of solids. Our examples will typically be dominated by shear rather than elongation. Nonetheless, phenomena such as ``extensional necking''~\cite{Fielding:2011} are analogous to what could be called ``elongation banding'', and many of the things mentioned in the review about shear-dominated flows could be carried over to elongation-dominated flows.  
\section{Shear banding in polymeric fluids\label{polym}}
In this section, we discuss shear banding in polymeric fluids, chiefly polymer and surfactant wormlike micelles solutions. We focus on the semi-dilute and concentrated regimes where polymeric chains form a viscoelastic entangled network. At or close to equilibrium, polymer and micellar solutions present a formal analogy~: they follow simple scaling laws as a function of concentration and they exhibit similar phase behaviors. The main discrepancy lies in the ability of wormlike micelles to break and recombine, a particularity that earned them the name of ``living polymers''. The analogy can be taken further when considering the rheology of these systems, especially because the original models able to capture much of their rheological features all originated from the tube model~\cite{Doi:1988}. 

\begin{figure*}
 \includegraphics[width=6in]{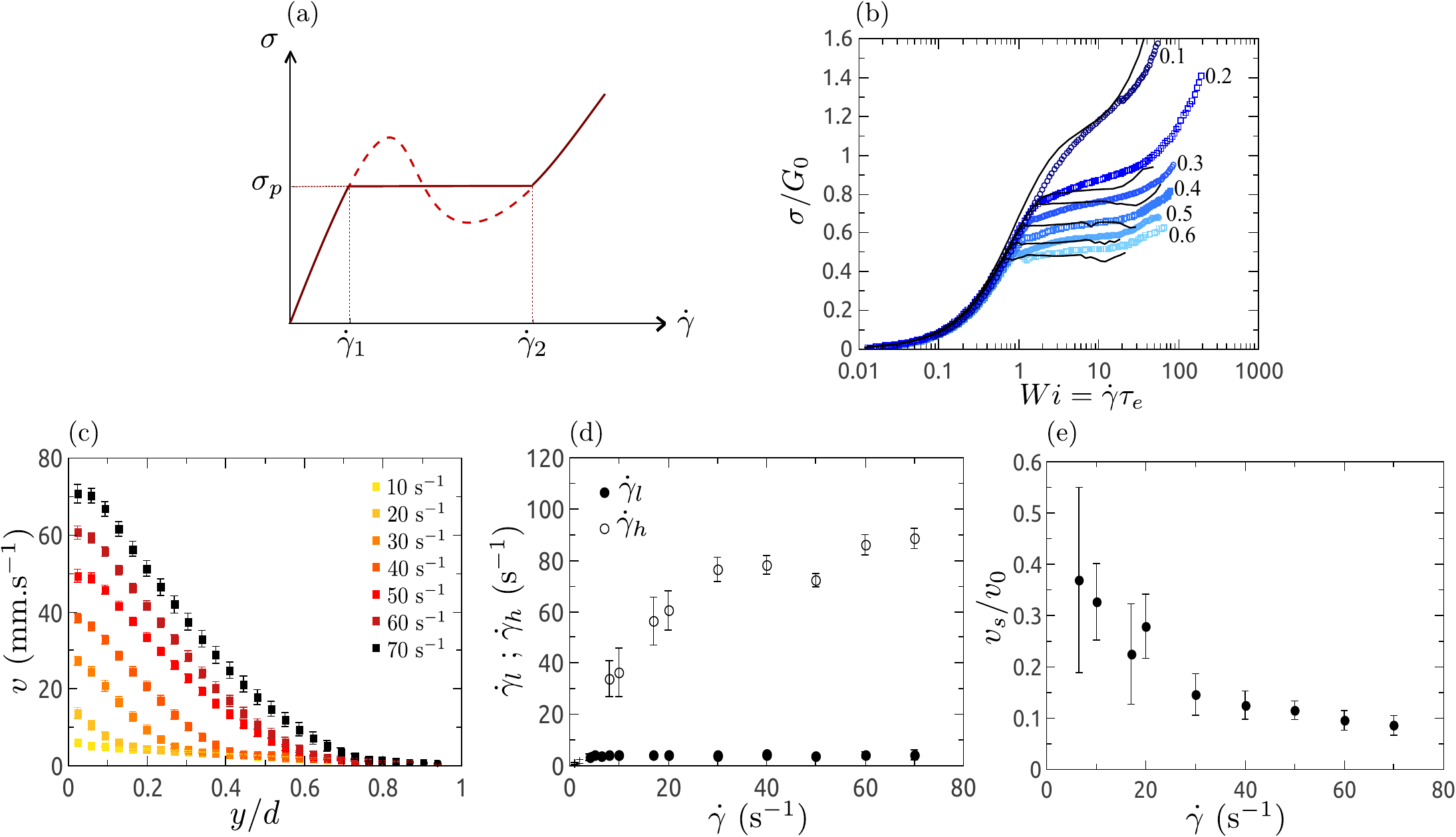}
 \caption{(a) Schematic flow curve of a complex fluid undergoing a shear banding transition according to the ``material instability scenario''. The measured flow curve is made
 	of two branches, \textit{a priori} stable, separated by a stress plateau ($\sigma=\sigma_p$). The underlying constitutive curve is non-monotonic. (b) Typical set of experimental steady-state flow curves of wormlike micelle solutions obtained for various surfactant concentrations by increasing the imposed shear rate. The shear
 	stress $\sigma$ is in units of the elastic modulus $G_0$ while the dimensionless shear rate is the	Weissenberg number $Wi\equiv\dot{\gamma\tau_e}$ where $\tau_e$ is the relaxation time of the fluid. Symbols and lines correspond to flow curves obtained respectively in a Taylor-Couette device and in cone-and-plate geometry. Figure adapted with permission from Fardin \textit{et al}, \textit{Soft Matter}, 8, 10072, 2012. Copyright by the Royal Society of Chemistry. (c) Typical time-averaged velocity profiles gathered in TC flow using ultrasonic velocimetry for different applied shear rates along the stress plateau. (d) Mean local shear rates $\dot{\gamma_l}$ and $\dot{\gamma_h}$ estimated from a linear fit of the velocity profile $v(y)$ in each band as a function of the applied shear rate. (e) Mean slip velocity $v_s$ measured at the inner moving cylinder as a function of $\dot\gamma$. Figures adapted  with permission from Fardin \textit{et al}, \textit{Eur. Phys. J. E}, 35, 1, 2012. Copyright by Springer-Verlag.}
 \label{fig1}
 \end{figure*}

Historically, shear banding has been first predicted and/or observed in this class of complex fluids which still constitutes a strong guide for the emergence of this phenomenon in other types of systems. In the context of solid mechanics, shear bands have been understood as a consequence of a ``material instability''~\cite{Goddard:2003}, which results --at the macroscopic continuum scale-- from non-monotonic constitutive relations. For fluids, constitutive laws are first and foremost probed by measuring the flow curve, i.e. the relationship between shear stress and shear rate in viscometric flows. If a fluid displays a non-monotonic flow curve as the one sketched in \textbf{Figure \ref{fig1}\textit{a}}, the range of shear rate with decreasing shear stress is mechanically unstable~\cite{Yerushalmi:1970}. Such curve is evidently not found in experiment, but computed theoretically by enforcing the homogeneity of the shear rate. In practice, the stress plateaus to a value $\sigma_p$ above a critical shear rate $\gp_1$, and the flow splits in domains bearing high ($\gp_h$) or low ($\gp_l$) shear rates, the low and high local shear rate being connected to the low ($\gp_1$) and high bounds ($\gp_2$) of the global stress plateau. In simple shear flows, the viscosities in the high and low shear bands are $\eta_h\sim\sigma_p/\gp_h$ and $\eta_l\sim\sigma_p/\gp_l$. That the local viscosities are changing is a testimony of the fact that the structure of the fluid is changing. In this context, the shear banding instability is often associated with shear-induced structuration of the fluid. 
The possibility for such material instability in fluids first emerged in research by~\cite{Vinogradov:1973}, on the ``spurt'' effect in polymer solutions and melts, \textit{i.e.} the dramatic increase of the flow rate above a critical pressure drop during extrusion out of a conduit. The idea that the ``spurt'' effect could be a consequence of the existence of shear banding was later formalized in particular by~\cite{McLeish:1986} who highlighted the fact that the then recently developed Doi-Edwards (tube) theory of polymer dynamics did predict a non-monotonic flow curve~\cite{Doi:1988}.  The problem with this hypothesis was that the many experiments on polymers done in viscometric conditions had not reported any non-monotonic or discontinuous flow curve, and a fortiori no shear banding~\cite{DPL}. From this discrepancy between theory and experiments, two routes were taken in parallel. On the one side, additions were made to the tube theory to take into account the kinetics of breaking/recombination in wormlike micelles~\cite{Cates:1990}, where broad stress plateaus hinting at the presence of shear banding were beginning to be observed in experiments~\cite{Rehage:1991}. This was the starting point of intensive experimental research sustained by a strong theoretical feedback. On the other side, the tube model was refined to better reflect the dynamics of polymers, and to ``cure'' the original model from its non-monotonicity~\cite{Marrucci:1996}. In parallel, experiments were conducted to support (or not) the theoretical  picture which was also updated regularly.

In what follows, we summarize the major advances in the field over the past ten years. We first discuss the successes and difficulties in correlating extensive sets of data in shear banding wormlike micelles, providing a roadmap for future investigations on other complex fluids. We then turn to polymer solutions and melts for which the existence of shear banding is still a very controversial subject. Finally, we give an overview on theoretical modeling of shear banding in polymeric fluids at large.  

\subsection{Correlating data sets of increasing subtlety in wormlike micelles~: a roadmap for other complex fluids}
\label{roadmap}
The non-monotonic flow curve sketched in Figure~\ref{fig1}\textit{a} is a powerful picture to understand shear banding; it is very general and applies to many types of flows and fluids. On this basis, purely rheometric experiments showed that for a given solution of wormlike micelles at a given temperature, the stress plateau was unique but could be influenced by the flow geometry. This became an important theoretical problem, which we will recount in section~\ref{theo}. However, this sketch is essentially zero-dimensional (0D) in the sense that it is obtained by imposing the tensorial strain rate field to be a pure shear rate field, constant and equal to the global shear rate signal $\bar\gp$. Space has disappeared from the picture, and since the flow curve is assumed to be steady, time is not mentioned either. 

It became quickly obvious that one had to go beyond this 0D picture and a collective effort was made to combine purely rheometric data with more mesoscopic, local and non-mechanical data (chiefly optical and structural data). Direct evidence for shear banded flows associated with the coexistence of differing micro-structures  was then supported by flow birefringence (FB), small angle neutron scattering (SANS) and nuclear magnetic resonance (NMR) velocimetry and spectroscopy experiments. For exhaustive references about that part of the story, the reader can refer to~\cite{Berret:2006,Lerouge:2010}. Heterogeneous flows have been probed in various flow geometries such as pipe~\cite{Mair:1997}, cone-and-plate (CP)~\cite{Britton:1997}, Taylor-Couette (TC)~\cite{Salmon:2003} and more recently  microchannels~\cite{Nghe:2010,Ober:2011}. Our understanding of shear banded flows in wormlike micelles is particularly advanced in TC flow, in the small gap limit, especially because this canonical geometry turned out to be the most suitable for the implementation of powerful space- and time-resolved techniques~\cite{Manneville:2008,Callaghan:2008,Eberle:2012}. Besides it presents additional advantages~: the slight stress inhomogeneity due to curvature tends to produce a plateau with a slight positive incline (\textbf{Figure~\ref{fig1}\textit{b}}), such that both stress and shear rate can be used as control parameters. It also pushes the shear banding flow to be made of only two bands, with the high shear rate band always near the inner cylinder (radius R). In contrast, in a CP flow, there can be two, three or even more bands, and their location is more variable~\cite{Britton:1997,Boukany:2008,Casanellas:2015}. TC flow is less prone to edge effects which can strongly limit the accessible range of flow strength in CP for instance~\cite{Dimitriou:2012}.
\begin{figure*}
	\includegraphics[width=6in]{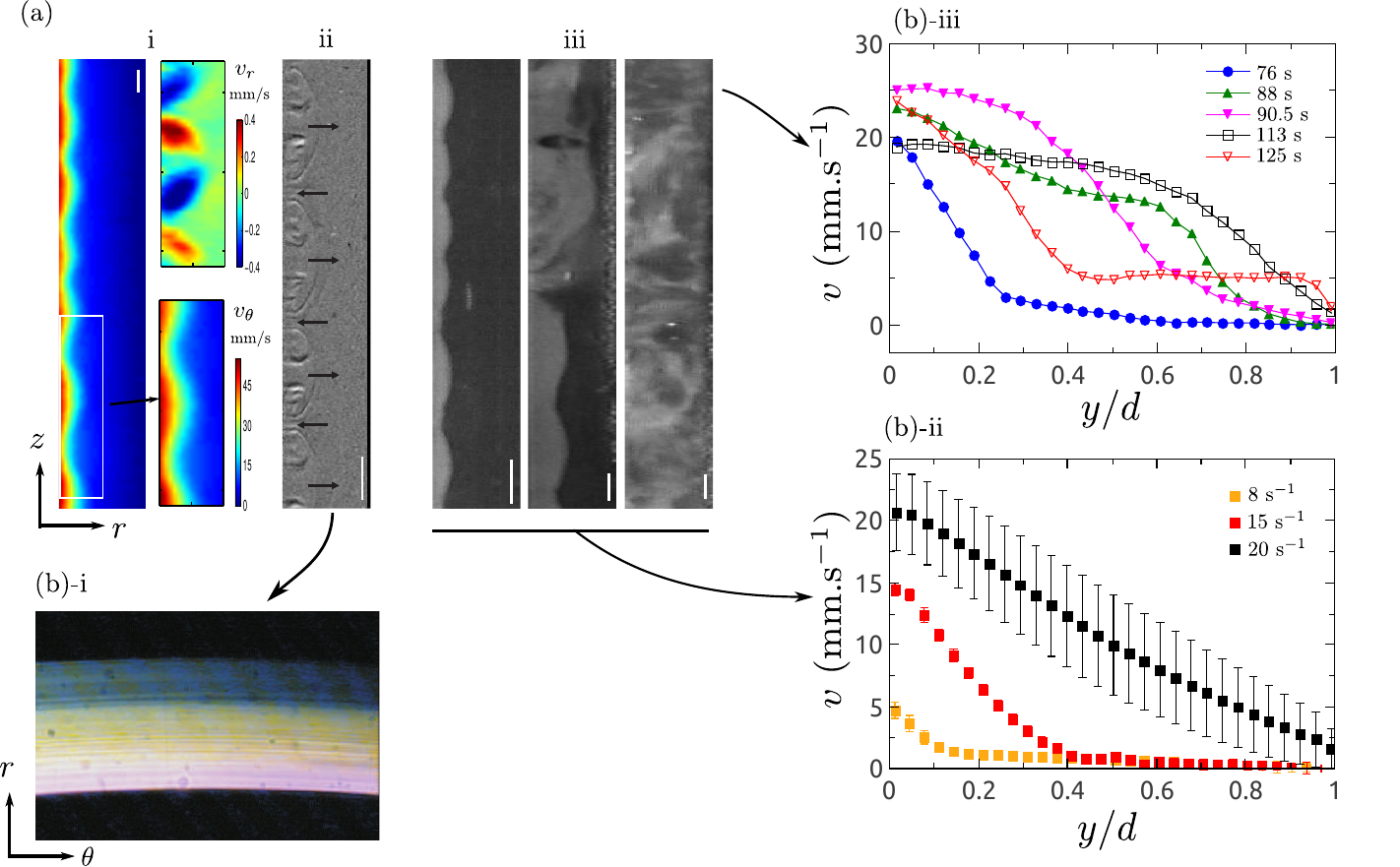}
	\caption{(a) 2D picture of shear banding flow in semi-dilute wormlike micelles illustrating evidence for elastic instability and turbulence in TC flow. i) Steady-state velocity map $v_{\theta}(r,z)$ deduced from ultrasonic imaging for a given shear rate. The gap size (2~mm) corresponds to the width of the picture while the bar (2~mm) gives the vertical scale. The maps at the right are enlargements showing respectively the radial $v_{r}(r,z)$ and azimuthal $v_{\theta}(r,z)$ velocity components. ii-iii) Views of the gap in the $(r, z)$ plane illuminated with either white light or laser light, the former providing direct visualisation of the vortex structure responsible for the undulation of the interface. The gap size (1.13~mm) corresponds to the width of the picture while the bar (1 mm) gives the vertical scale. In i) and ii) the sytem is made of cetyltrimethylammonium bromide and sodium nitrate while iii) corresponds to a sample of cetylpyridinium chloride with sodium salycilate in brine for different applied shear rates.  Figure adapted with permission from Fardin \textit{et al}, \textit{Soft Matter}, 10, 8789, 2014. Copyright by the Royal Society of Chemistry. (b) 1D picture of shear banding flow in semi-dilute wormlike micelles. i) Flow birefringence snapshot in the ($r,\theta$) plane. The annular gap is illuminated with white light and placed between crossed polarizers. The evidence of three bands is purely artefactual. The intermediate band results from the interfacial undulation and leads to smooth orientation profiles between bands. Figure adapted with permission from Lerouge \textit{et al}, \textit{Langmuir}, 20, 11355, 2004. Copyright by the American Chemical Society.  ii) Time-averaged velocity profiles for different applied shear rates associated with pictures (a-iii). iii) Selection of several instantaneous velocity profiles illustrating the impact of a turbulent burst on the main flow. Figure adapted with permission from Fardin \textit{et al}, \textit{Soft Matter}, 8, 10072, 2012. Copyright by the Royal Society of Chemistry.}
	\label{fig2}
\end{figure*}
Typical time-averaged velocity profiles gathered in TC flow are displayed in \textbf{Figure~\ref{fig1}\textit{c}} as a function of the radial distance $y$ to the inner cylinder normalized by the gap width $d$. Such velocity profiles now provide what we could call a 1D picture. The early 1D experiments~\cite{Salmon:2003} were mostly focused on the simple assumptions made in theoretical modeling~: no wall slip, infinitely sharp interface, and lever rule (local shear rates equal to global bounds of the stress plateau and linear increase of the proportion of the high-shear band with the global shear rate)~\cite{Spenley:1996}, which constitute the so-called ``classical scenario''. With the refinement of spatial and temporal accuracy, we now understand these assumptions to be simplistic. Wall slip on the high shear band seems quite deeply connected to shear banding, and leads to deviations from the original lever rule~\cite{Radulescu:1999,Fardin:2012}. In particular, the local high shear rate is usually observed to be somewhat in competition with wall slip~\cite{Lettinga:2009,Feindel:2010,Fardin:2012a}. It increases with the applied shear rate, as the dimensionless wall slip is concomitantly reduced, as shown in \textbf{Figures~\ref{fig1}\textit{d,e}}. This feature emerges even in the simplest theories of shear banding due to the coupling between the diffusive terms and the boundary conditions~\cite{Cromer:2011,Fardin:2012}. As for the width of the interface $\ell$, it is thin in comparison to usual gap ($d\simeq 1$~mm), but it is larger than what was once thought~\cite{Radulescu:2003}, and is typically measured to be ranging between 1 and 10 microns~\cite{Ballesta:2007}. In microfluidic devices, the existence of such a length scale leads to so-called ``non-local effects''~\cite{Masselon:2010}. This 1D picture has also been developed on the structural side. Orientation profiles were established using FB~\cite{Lerouge:2004}, NMR spectroscopy~\cite{Lopez:2006} and more recently SANS~\cite{Helgeson:2009,Gurnon:2014}. They all showed consistency with flow-aligned wormlike micelles in the high shear rate band coexisting with entangled wormlike micelles in the low shear rate band for semi-dilute samples and with  shear-induced isotropic-to-nematic transition for concentrated samples, with possibly concentration differences between bands on the latter case.

With the improvement of time resolution, local velocity fluctuations and complex motions of the bands have been reported in many different wormlike micellar systems, often coupled to non-
stationary responses of the global rheological signals~\cite{Lerouge:2000,Lopez:2004,Hu:2005,Ganapathy:2006,Becu:2007,Fielding:2007}. Understanding of these complex fluctuating dynamics recently came from extension of the experiments to 2D, 2D+time, etc. Indeed, in the 1D picture, the velocity component in the flow direction is obtained at one given height along the cylinders while FB and SANS data are gathered averaging along the vorticity direction. These specifications are now seen as quite limiting, and sometimes even misleading. The fluctuations in shear banding wormlike micelles were shown to be mainly due to the development, in the high shear rate band, of secondary flows either coherent and associated with interfacial undulations along the vorticity direction~\cite{Lerouge:2008} or turbulent~\cite{Fardin:2009,Fardin:2012a,Perge:2014a} (\textbf{Figure~\ref{fig2}\textit{a}}). The impact of such 3D shear banding flows on the 1D picture was fully characterized (\textbf{Figure~\ref{fig2}\textit{b}}), demonstrating the ubiquity of this phenomenon~\cite{Fardin:2012c}, which was shown to originate from viscoelastic instabilities driven by normal stresses akin to those well-known to develop in polymer solutions in flows with curved streamlines~\cite{Morozov:2007,Muller:2008}. The most recent experiments and simulations have suggested that in curved shear banded flows, bulk and interfacial viscoelastic instabilities can be at play, the bulk mechanism being dominant except when the effective curvature of the high shear band is vanishing~\cite{Fielding:2010,Decruppe:2010,Fardin:2011,Nicolas:2012}. These recent advances may provide a strong guide for the understanding of complex shear banding pictures in other flow geometries. 
\subsection{Shear banding in polymers solutions and melts}
As mentioned in the preamble of section~\ref{polym}, despite the successes of the tube theory to capture much of the rheology of entangled polymer solutions and melts, for a long time shear banding was not reported in these systems~\cite{Menezes:1982,Bercea:1993,Pattamaprom:2001}. All the early studies were purely rheometric in echo to the purely ``mechanical instability'' illustrated by the 0D picture. As shear banding was not observed, theoreticians worked on updating the original tube model (see~\cite{McLeish:2002} for a review). One of the additions particularly relevant to shear banding is convective constraint release (CCR), which enables the relaxation of some stress carried by the test chain due to the reconfiguration of the tube when an entanglement point is lost as chains slide past each other. Such CCR becomes increasingly important with shear rate and was shown to be able to remove the non-monotonicity of the original tube model. Note that arguments were put forth to justify a lesser degree of CCR in wormlike micelles (due to breaking/recombination) and thus allowing the existence  of shear banding in micelles~\cite{Milner:2001}. These arguments were shown correct in very recent simulations~\cite{Zou:2014}. 
\begin{figure*}
	\includegraphics[width=6in]{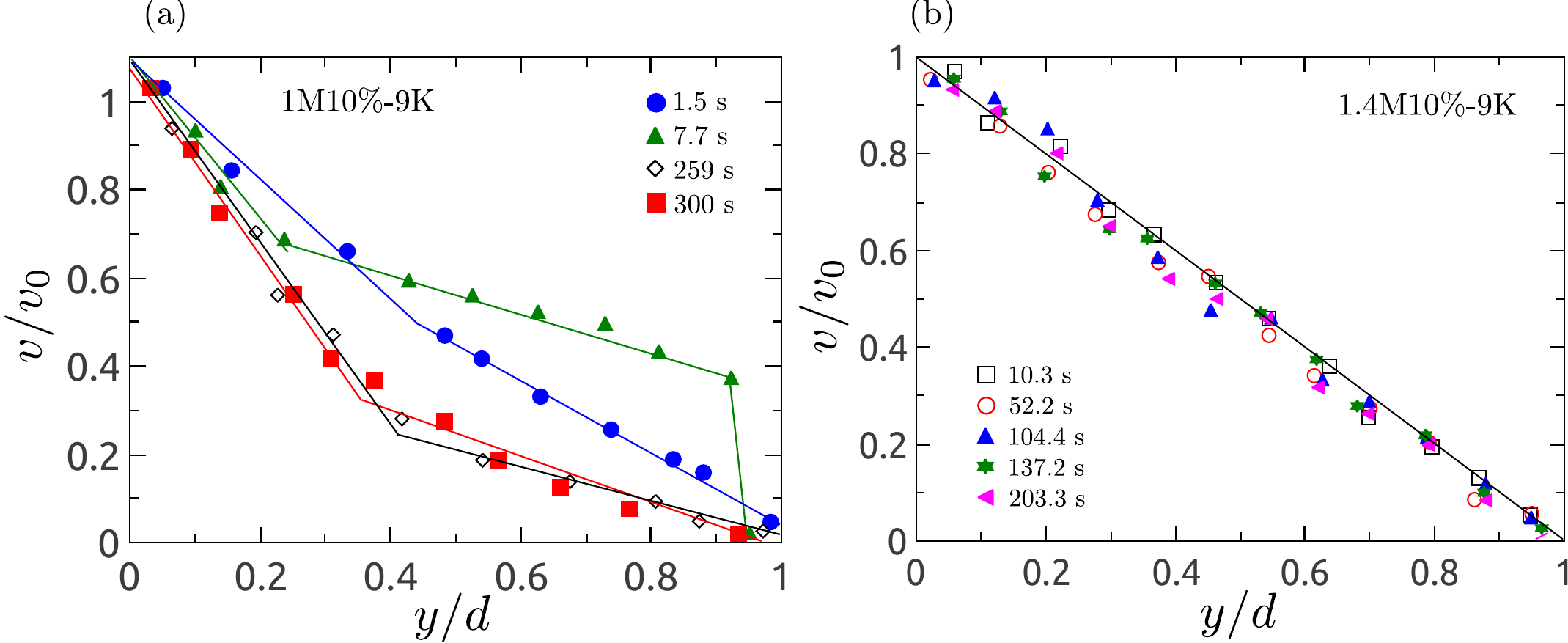}
	\caption{(a) Velocity profiles at different times following startup of flow and gathered in rotating parallel plate geometry. $H_0$ is the gap size and $v_0$ is the velocity of the moving plate at the radius where local velocities are measured. The sample is a polybutadiene solution (1M10\%-9K) with an entanglement density $Z=42$. Figure adapted with permission from Ravindranath \textit{et al}, \textit{Macromolecules}, 41, 2663, 2008. Copyright by the American Chemical Society. (b) Velocity profiles following a startup shear in parallel plate geometry. The sample is a polybutadiene solution (1.4M10\%-9K) with an entanglement density $Z=55$. Figure adapted with permission from Li \textit{et al}, \textit{J. Rheol.}, 57, 1411, 2013. Copyright by the Society of Rheology. }
	\label{fig3}
\end{figure*}
Since 2006, strong efforts have been made to provide a 1D picture using velocimetry techniques in various flow geometries, mostly CP~\cite{Ravindranath:2008,Li:2013} and parallel plates~\cite{Boukany:2007,Hayes:2008} and marginally TC~\cite{Hu:2010} and pipe~\cite{Wang:2013}. During the past decade, extensive sets of particle tracking velocimetry (PTV) experiments combined with global rheology by Wang's group~\cite{Wang:2011} showed that shear banding can emerge in polymer solutions and melts (\textbf{Figure \ref{fig3}\textit{a}}). Recently, shear banding in polymers has also been evidenced by other groups using optical coherence tomography~\cite{Jaradat:2012} or PTV~\cite{Noirez:2009}. As in wormlike micelles, shear banding was often found to compete with wall slip~\cite{Boukany:2008,Jaradat:2012} which seems ubiquitous in these systems and can be limited by surface roughness and/or treatment and use of entangled polymeric solvent. The emergence of shear banding in polymer solutions and melts seems to require sufficiently entangled~\cite{Boukany:2009a} and monodisperse~\cite{Boukany:2007} samples. Furthermore the competition between shear banding and wall slip seems to depend both on the concentration and molecular weight~\cite{Wang:2011,Jaradat:2012} and bulk banded flows are likely to overcome slip effects for sufficiently high applied shear rates~\cite{Boukany:2008,Boukany:2009}. Finally, the existence of steady shear banding seems connected to the flow history, at least for  solutions~\cite{Boukany:2010,Wang:2012}. Indeed, shear banding seems to be observed at steady state following startup shear protocols while it seems to be a transient property when using shear ramp protocols. 

However, up to now, the 1D picture described above has not reached a consensus. PTV experiments in TC flow showing the possible existence of both banded and non-banded profiles for a given sample in the supposed regime of shear banding suggested that shear banding may be only a long-lived transient feature in polymer solutions~\cite{Hu:2010}. In contrast, other groups were unable to observe shear banding neither transient nor steady~\cite{Hayes:2008,Hayes:2010,Li:2013} in systems close to those investigated by Wang's group (\textbf{Figure \ref{fig3}\textit{b}}). They concluded that wall slip and linear velocity profiles prevail in the response of polymer solutions. In addition, it was suggested that edge effects and/or experimental artifacts may be responsible for the observed shear banding in polymer solutions~\cite{Li:2013,Li:2014}. 

This situation gave rise to impassioned debates~\cite{Adams:2009,Wang:2009,Adams:2009b}. Wang's interpretation of shear banding involves microscopic failure occurring within the bulk under large step shear, thus challenging the tube picture which cannot account for such ``elastic yielding''. Hence, rejecting almost the entire reptation picture on one side, and casting doubt on the quality of experiments on the other, the controversy on shear banding in polymers is far from being sorted out~\cite{Wang:2014,Li:2014}. Nonetheless, it stimulated much theoretical work trying to provide rationales for the various experimental observations, as illustrated in the next section. 
\subsection{Theoretical frameworks for shear banding in polymeric fluids\label{theo}}
We have already mentioned that the tube-like models with all their modern additions can predict shear banding in entangled polymeric fluids~\cite{Milner:2001} and wormlike micelles~\cite{Cates:2006}. However these microscopic models are not always easily tractable in the nonlinear flow regime. In practice more ``phenomenological models'' are used instead~\cite{Cates:2006,Olmsted:2008}. These models do not usually contain all the information on the dynamics of the microstructure and deal with coarse-grained quantities defined at the macroscopic scale. They usually at least include tensorial stress and velocity gradient fields. Simple phenomenological models like the diffuse Johnson-Segalman model (d-JS) or the diffusive Giesekus model (d-Giesekus) can qualitatively reproduce many of the macroscopic aspects of shear banding. Nonetheless, they are too coarse-grained to differentiate between micelles and polymers. A compromise between microscopic and phenomenological models is the so-called ``Rolie-Poly'' model, which is a simplified differential version of the tube model that incorporates CCR. 

Phenomenological models of shear banding in polymeric fluids usually rely on a disjunction of the stress in two parts, a ``solvent part'' ($\eta_s\gp$), and a ``polymer part'' ($\eta_p\gp$). The polymer part, like in the tube models is only increasing up to a critical shear rate, after which it continuously decreases. Ultimately, the viscosity at infinite shear rate is $\eta_s$. Shear banding emerges if the so-called ``viscosity ratio'' ($\eta\equiv\eta_s/\eta_p$) is small enough for the total flow curve to be non-monotonic. This behavior is also observed experimentally in wormlike micelles, by varying the concentration in surfactant or the temperature, as shown in Fig.~\ref{fig1}b. Above a critical viscosity ratio, $\eta_c$, steady shear banding disappears.       
Furthermore, to account for experimental observations in wormlike micelles of a unique stress plateau at $\sigma_p$ independent of the shear history and the initial conditions, non-local (diffusive) terms have been included in the equation governing the polymeric stress (see~\cite{Olmsted:2008} for a review). Such ``diffusion'' terms are now widespread, and allow one to define the typical width of the interface between bands as a function of the stress diffusion coefficient, $\ell\equiv\sqrt{\mathcal{D} \tau_e}$ where $\tau_e$ is the relaxation time of the fluid. During the past decade, phenomenological models including diffusive terms were used to test the impact of the flow geometry~\cite{Radulescu:2000} and the boundary conditions~\cite{Adams:2008} on the banding structure, the role of a flow-concentration coupling~\cite{Fielding:2003b}, the effect of the control parameter~\cite{Dhont:1999}, or more recently, the stability of the shear banding flow~\cite{Fielding:2010,Nicolas:2012}. 

In tube models and in the more phenomenological models like d-JS, the shear banding instability is seen from the mechanical perspective we mostly echoed so far. Shear banding in wormlike micelles can also be approached from a more ``thermodynamic perspective'', noting the similarities between shear banding and first-order phase transitions. For instance, the critical viscosity ratio $\eta_c$, leads to a flow curve with an inflection point that shares much similarity with a critical point. Unfortunately, this critical point as remained largely unexplored experimentally. Generally, Fig.~\ref{fig1}b can be called a ``flow-phase diagram''. If the thermodynamic and mechanical perspectives were once competing, they are now understood as the two sides of the same coin.        
Let us note that if for polymeric fluids the rationale for shear banding provided by the tube picture seems to be the correct one, other possibilities are nonetheless investigated. In particular, in wormlike micelles, the Vasquez-Cook-McKinley (VCM) model has been shown to be able to reproduce many of the properties of shear banding observed in experiments~\cite{,Zhou:2014}. This model devised at the intermediate level of kinetic theories is built out of two species of dumbbells, short and long, where two short ones are allowed to combine into a long one, and with long ones increasingly breaking apart as the flow strength is increased. In effect, shear banding in this model is a consequence of shear-induced demixing of long and short species, with long ones eventually disappearing entirely. This explanation for shear banding differs from the one provided by tube models, and whether or not it applies to shear banding in wormlike micelles should be checked experimentally by investigating if the typical length of micelles at high shear rate is indeed shorter than at low shear rate. 

Finally let us come back to the  controversial situation in polymer solutions and melts. Using the diffusive ``Rolie-Poly'' model, Adams and Olmsted first highlighted that transient shear banding could appear in polymeric flows for viscosity ratios $\eta$ just above the critical value $\eta_c$, i.e. for flow curves that are actually monotonic or equivalently for nonbanded steady states~\cite{Adams:2009,Adams:2011}. Key ingredients for the emergence of transient shear banding are thermal noise or high stress gradients inherent to the flow geometry. This observation was subsequently extended by Moorcroft and Fielding, who derived a very general mechanical criterion encompassing all complex fluids to predict the occurrence of transient as well as steady shear banding~\cite{Moorcroft:2011,Moorcroft:2013}. Checking the validity of this criterion for wormlike micelles solutions with viscosity ratio above and below the limit $\eta_c$ appears as a logical next step for experiments. The diffusive Rolie-Poly model was also recently used to offer further ways to interpret the data on shear banding in polymers~\cite{Olmsted:2013}. Recently, another alternative involving the Rolie-Poly model was also shown to reproduce qualitatively almost all the phenomena observed by Wang's group for a model polymer solution with an underlying monotonic constitutive curve, the driving mechanism being the coupling of the polymer stress to an inhomogeneous concentration profile~\cite{Cromer:2014}.

%%%%%%%%%%%%%%%%%%%%%%%%%%%
\section{Shear banding in soft glassy materials}
\label{softglass}
%%%%%%%%%%%%%%%%%%%%%%%%%%%

Having reviewed the current state of knowledge on shear banding in polymeric solutions, we now discuss shear banding in Soft Glassy Materials (SGM). Here we shall take the term ``glassy'' in a broader sense than generally assumed, by encompassing complex fluids that can either be liquid-like or solid-like at rest but that are all characterized by a microstructure where ``elementary bricks'' strongly interact, e.g. through steric constraints, entanglements, attractive forces or physical bonds. With this in mind, liquid-like SGM correspond to solutions of surfactants or copolymers yet at  higher concentrations than the wormlike systems discussed so far. These solutions display highly viscoelastic lyotropic mesophases such as hexagonal, lamellar, cubic and bicontinuous ``sponge'' phases. On the other hand solid-like SGM are systems which micro-constituents are kinetically trapped into a disordered, metastable configuration and where thermal energy alone is not sufficient to significantly rearrange the microstructure and relax mechanical stresses. The glassy behavior of such SGM originates from either geometric frustration or attractive interactions between the constituents and they typically exhibit a ``yield stress,'' i.e. they can rearrange and flow provided a stress larger than some critical value $\sigma_c$ is applied. Examples of solid-like SGM range from concentrated jammed systems of repulsive hard spheres, soft particles or liquid droplets to dispersions of attractive colloids that form space-spanning networks even at very low volume fractions, i.e. colloidal gels.

\begin{figure*}
\includegraphics[width=\linewidth]{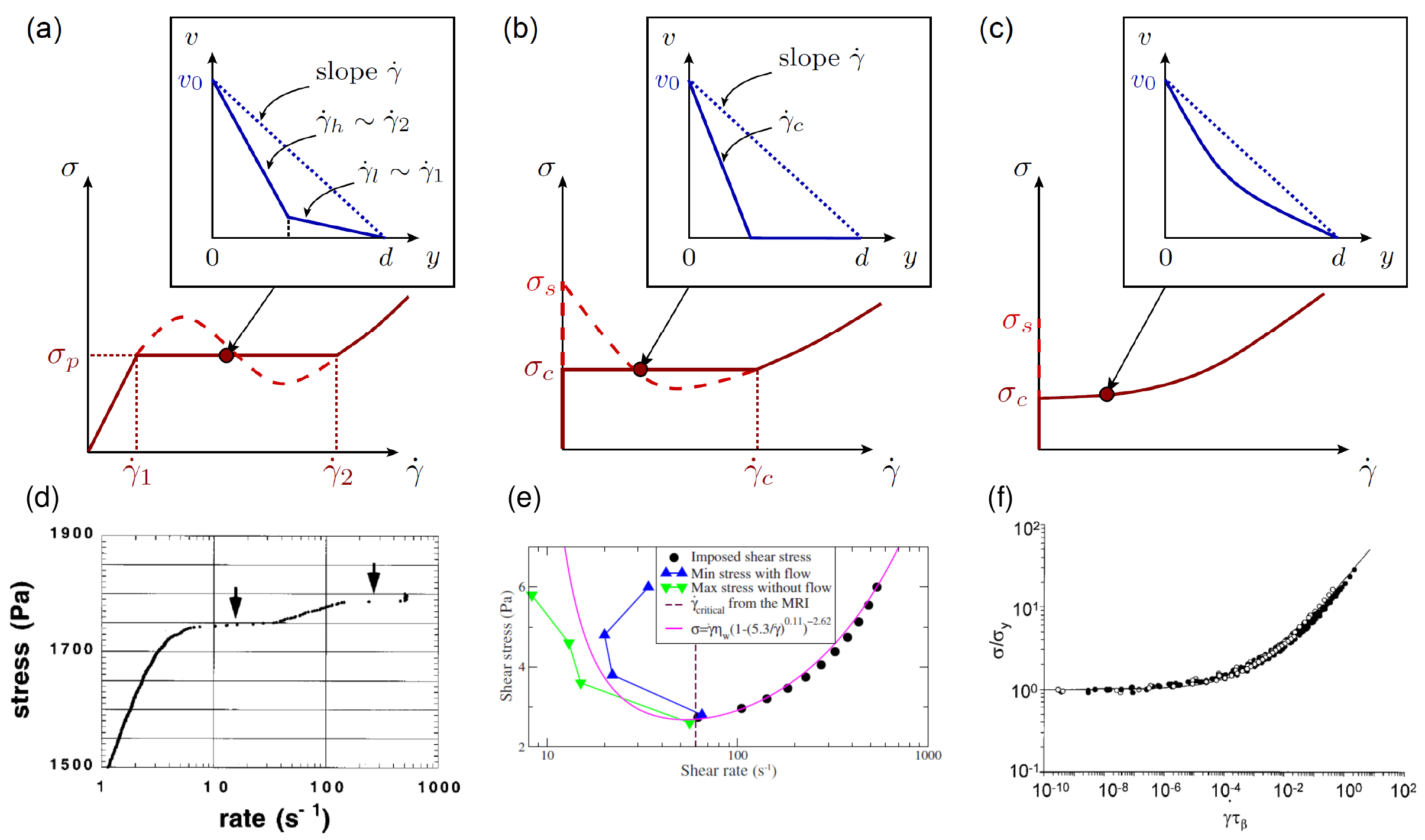}
\caption{Sketches of the flow curves $\sigma$ vs $\dot \gamma$, and velocity profiles $v(y)$ expected in (a)~materials undergoing a mechanical instability and/or a shear-induced transition such as semi-dilute wormlike micelle solutions or lamellar and cubic phases, (b)~a yield stress material with steady-state shear banding below a critical shear rate $\gp_c$ and (c)~a yield stress material that follows a Herschel-Bulkley rheology and flows homogeneously in steady state. $\sigma_p$ stands for the value of the stress plateau in (a) while $\sigma_c$ and $\sigma_s$ denotes respectively the apparent yield stress and the static yield stress in (b) and (c) (see text). In the sketched velocity profiles, the blue dotted lines indicate the case of a Newtonian fluid. $v_0$ stands for the velocity of the moving wall. These three different types of flow curves are illustrated for (d)~a cubic phase of a triblock copolymer made of ethylene and propylene oxide showing two successive stress plateaus, (e)~a colloidal suspension of Ludox silica spheres and (f)~microgels with different crosslink density made of acrylate chains bearing methacrylic acid units. In the latter case, the flow curve has been rescaled by the yield stress and the fluid relaxation time $\tau_\beta$. The continuous line is the best Herschel-Bulkley fit. (d), (e) and (f) are adapted with permission from respectively \cite{Eiser:2000}, \cite{Moller:2008} and \cite{Cloitre:2003}. Copyrighted by the American Physical Society.
}
\label{fig4}
\end{figure*}

There is currently no universal explanation for heterogeneous flows in SGM, and when compared to wormlike micellar fluids, the picture is at best fragmented if not controversial. Clearly, difficulties in characterizing and understanding shear banding in SGM arise from the fact that, as the zero-shear viscosity increases and eventually goes to infinity upon jamming or gelation, relaxation times considerably increase so that at least one of the characteristic shear rates $\gp_1$ and $\gp_2$ apparently goes to zero as sketched in \textbf{Figure~\ref{fig4}}. In particular, in order to be conclusive, any study involving low shear rates and/or focusing on solid--fluid coexistence must be performed over long enough timescales to ascertain that a steady state is reached. Over the years, it has progressively been recognized that, depending on the nature of the fluid microstructure and on the interactions between its constituents, shear banding in SGM originates from one or a combination of the following causes: ($i$) an underlying shear-induced phase transition, ($ii$) the competition between shear and the attractive interactions between the constituents or ($iii$) flow--concentration coupling. For instance, a dilute assembly of monodisperse colloidal hard-spheres may develop a shear band due to flow-induced crystallization \cite{Shereda:2010}, while more concentrated and slightly polydisperse samples may experience the formation of a permanently arrested band due to minute variations of the local packing fraction that trigger the jamming of a macroscopic region of the flow \cite{Besseling:2010}. 
 
In what follows we have chosen to deepen the phenomenology of shear banding in SGM through a historical approach put into perspective by recent progress in modeling. We first discuss the case of concentrated surfactants and block copolymers, for which shear banding mainly results from shear-induced structuration, before turning to colloidal star polymers that display a delayed scenario for shear band formation. SGM with a yield stress are the topic of the second and third subsections, where we respectively review steady-state shear banding --the most studied case so far-- then transient shear banding --an emerging area in SGM.

\subsection{From shear-thinning fluids to glassy-yet-not-jammed materials}

Concentrated surfactants and block copolymers exhibit lyotropic mesostructures which properties under shear have been the topic of numerous studies. Close to an equilibrium phase transition, shear may indeed ``help'' the system cross the phase boundary. Sheared liquid crystalline phases may also structure to form novel out-of-equilibrium shear-induced structures (SIS) and are likely to involve shear banding \cite{Berni:2002}. Under moderate shear some lamellar phases were shown to organize into a novel structure of disordered close-compact multilamellar vesicles called ``onions'' \cite{Diat:1993b}. This disordered onion assembly itself undergoes a shear-thinning transition at larger shear where the final state is characterized by hexagonally-ordered onion layers sliding on top of each other along the velocity direction \cite{Roux:1993}. In the vicinity of this ``layering'' transition, velocimetry revealed that the nucleation and growth of such SIS is associated with the nucleation and growth of a shear band, and that the classical picture of shear banding detailed in section~\ref{roadmap}, including the lever rule, holds true \cite{Salmon:2003d}. However, numerical simulations of lamellar systems that reproduce shear banding show no evidence for a stress plateau, which suggests that if the shear band is robust, the lever rule may not be universal in these systems \cite{Xu:2006}. 

A quite similar situation has been reported in concentrated copolymer solutions forming spherical micelles that arrange into cubic phases. Under shear the initial disordered polycrystalline structure successively orients into different bcc crystals, each transition being signaled by a stress plateau [\textbf{Figure~\ref{fig4}\textit{d}}] where two structures coexist as evidenced by x-ray diffraction \cite{Eiser:2000}. For a review of the shear-induced states in block copolymers, see \cite{Hamley:2001}. Note also that such orientation transitions are reminiscent of those observed in colloidal crystals \cite{Chen:1994}. While block copolymers forming wormlike micellar structures have been shown to display shear banding and elastic instabilities reminiscent of surfactant wormlike systems \cite{Manneville:2007}, the local flow properties of more concentrated copolymers remain largely unexplored and should deserve more attention in future studies to fully understand these shear-induced transitions \cite{Eiser:2000b}. Another line of research concerns so-called ``transient networks'' made of e.g. reversibly crosslinked telechelic copolymers or proteins [see \cite{Ligoure:2013} for a review]: while these systems have long been known to ``fracture'' under shear, a phenomenon associated with a decrease of the stress with strain \cite{Skrzeszewska:2010}, the shear banding recently reported in a numerical model of transient network \cite{Billen:2015} raises the fundamental question of whether fractures could be seen as an extreme case of shear bands.

\begin{figure*}
\includegraphics[width=\linewidth]{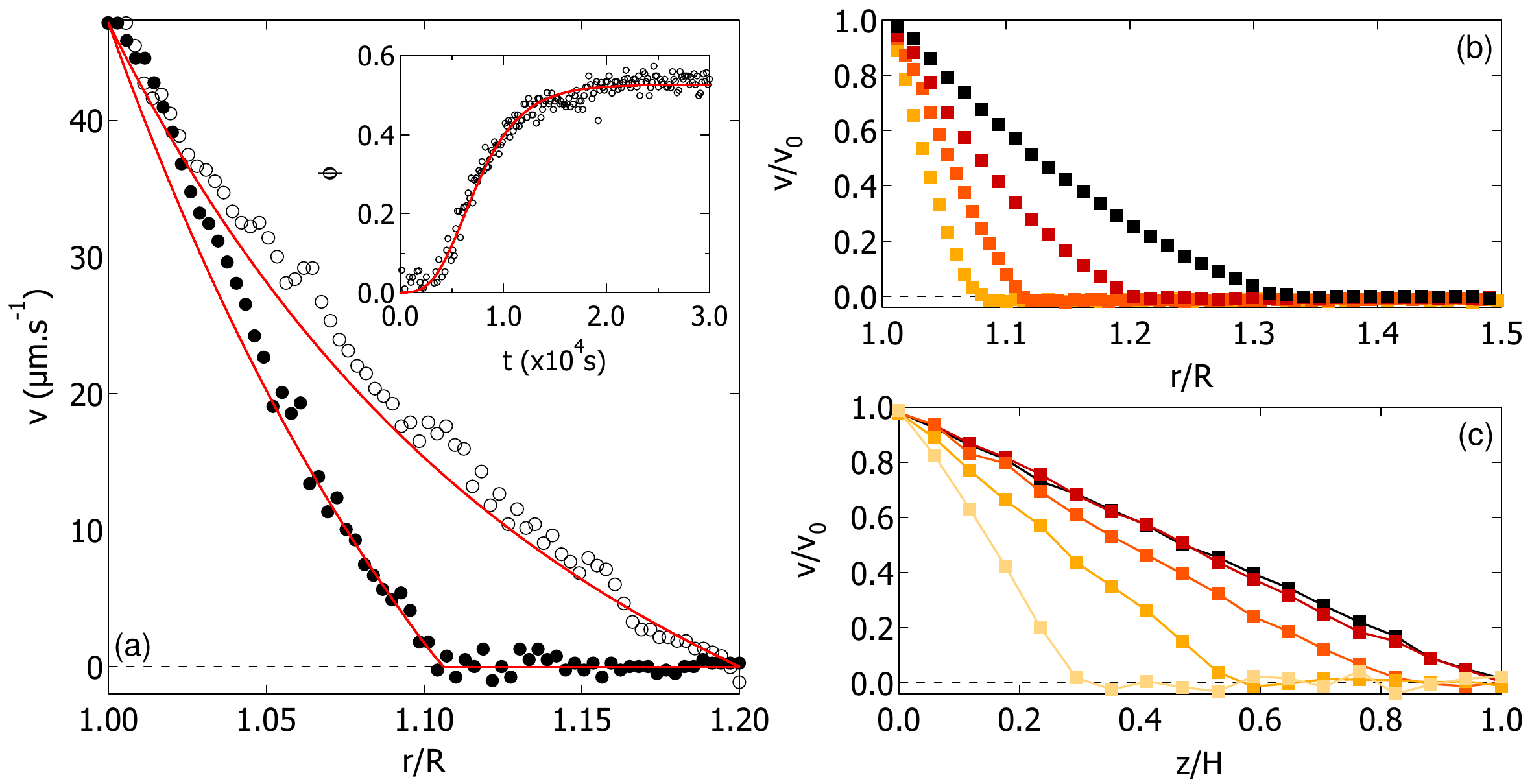}
\caption{(a) Velocity profiles at different times [($\circ$): $t=150$~s and ($\bullet$): $t=25000$~s] during a shear start-up experiment ($\dot \gamma =3.14\times 10^{-2}$~s$^{-1}$) for a system of colloidal star polymers sheared in a TC geometry of inner radius ${\rm R}=7.5$~mm. The red curves correspond to the predictions of a fluidity model. Inset: Temporal evolution of the fraction $\phi$ of jammed material in the gap, fitted by an exponential function (red line). Adapted with permission from \cite{Rogers:2008}. Copyrighted by the American Physical Society. (b) Normalized steady-state velocity profiles in a drilling mud in a wide gap TC geometry (${\rm R}=4$~cm), at different rotation speeds: $\Omega =4, 10, 20$ and $60$~rpm from left to right. %The fraction of flowing material increases linearly with $\Omega$, while the former flows at a constant shear-rate roughly independent of $\Omega$. (angle 4.5$^{\circ}$, diameter 12~cm)
Adapted with permission from \cite{Ragouilliaux:2006}. (c) Normalized steady-state velocity profiles in a TC geometry for a bentonite suspension at different rotation speeds: $\Omega=5$, 10, 20, 40 and 80~rpm. $z$ denotes the vertical position within the cone, while $H$ stands for the gap thickness ($H=4.7$~mm at the periphery of the cone). Adapted with permission from \cite{Ovarlez:2009}. Figures (b) and (c) are copyrighted by Springer-Verlag. 
}
\label{fig5}
\end{figure*}

Finally dense suspensions of hairy soft particles made of colloidal star polymers (CSP) display rheological properties intermediate between the above systems and the concentrated colloids addressed in the next section \cite{Christopoulou:2009,Vlassopoulos:2010}. They are liquid-like and shear-thinning at short times after an oscillatory pre-shear is applied but strikingly develop shear-banded velocity profiles over thousands of seconds as illustrated in \textbf{Figure~\ref{fig5}\textit{a}} \cite{Rogers:2008}. Such delayed transients towards a heterogeneous steady state are observed below a critical shear rate $\gp_c$ that depends on the particle concentration \cite{Holmes:2004,Rogers:2010}. 
The slow emergence of a yield stress and of a stress plateau spanning from 0 to $\gp_c$ in the flow curve of CSP suspensions has been attributed to the progressive entanglement of percolating groups of particles \cite{Rogers:2010} and similar long-lived transients have been reproduced in Brownian dynamics simulations \cite{vanderNoort:2008}. Such time dependence, or aging, is also key to the formation of steady heterogeneous flows as we shall also see below for SGM made of attractive colloidal particles.

\subsection{Steady-state shear banding in yield stress fluids}

In the early 1990's shear banding in yield stress fluids (YSF) was originally referred to as ``shear localization'' and denoted the coexistence of two (or more) partially sheared bands, often sliding along each other, during transient or steady-state flows indifferently. These heterogeneous flows had been identified thanks to seminal experiments conducted on complex materials often inherited from industry (ink, lubricating grease, etc.) in parallel plate and TC geometries. In a pioneering work, Piau and coworkers unveiled a wide range of heterogeneous flows in clay suspensions and carbopol microgels, including fractures, wall slip, coexistence of plug flow(s) and/or shear band(s) and stick-slip dynamics, by simply following a line painted along the vorticity direction at the sample periphery \cite{Magnin:1990,Pignon:1996}. 
In order to rationalize such a wealth of flows, experimentalists quickly focused on two other geometries that made flow visualization easier and more quantitative, namely TC geometry and channel flow. Coupled to local techniques such as light scattering, particle tracking, magnetic resonance, ultrasonic echography or confocal microscopy, rheological tests were much enriched by the simultaneous measurement of bulk velocity fields \cite{Manneville:2008,Callaghan:2008}. Historically, experiments in wide-gap TC cells have played a particular role in the understanding of shear banding. We recall that in TC geometry the local shear stress decreases from the inner to the outer cylinder as $1/r^2$, where $r={\rm R}+y$ denotes the distance to the rotation axis. Therefore, shear banding is somewhat trivially observed when the yield stress of the fluid under scrutiny lies between the maximum and minimum local stresses. Of course this situation also occurs in any channel flow as soon as the wall shear stress exceeds the yield stress, since the local stress always vanishes at the center of the channel. Such shear banding induced solely by the stress heterogeneity inherent to the geometry is now referred to as ``shear localization'' to make a clear distinction with the intrinsic shear banding related to the nonlinear rheology of the material \cite{Ovarlez:2009}, towards which we are now turning.

Interestingly, numerous YSF with strong attractive interactions display steady-state heterogeneous velocity profiles that cannot be attributed to the shearing geometry. The coexistence between a flowing band and a solid-like region rather results from the existence of a critical shear rate $\dot \gamma_c$ below which no steady homogeneous flow is possible \cite{Coussot:2002c,Coussot:2002a}. In terms of the flow curve, this corresponds to the case of wormlike micelles explored in section~\ref{polym} but where the lower characteristic shear rate $\gp_1$ would be set to zero and the upper limit of the stress plateau $\gp_2$ would correspond to $\dot \gamma_c$. This analogy, illustrated in \textbf{Figure~\ref{fig4}\textit{b,e}}, has been quantitatively pushed forward thanks to experiments performed on attractive colloids in TC geometry \cite{Coussot:2002a,Moller:2008,Ovarlez:2009}. For a homogeneous shear rate $0\leq \dot \gamma \leq \dot \gamma_c$, steady-state shear banding was observed with a flowing  band sheared at $\dot \gamma_c$ over a width roughly following the lever rule with $\gp_1=0$ and $\gp_2=\gp_c$ [\textbf{Figure~\ref{fig5}\textit{b,c}}]. However, the situation is far less clear than for wormlike micelles as ($i$) there is a lack of measurements on other systems and ($ii$) the chemical and physical nature of the boundary conditions as well as the preshear protocol prior to the experiment may strongly influence the results \cite{Gibaud:2008,Cheddadi:2012}. Last but not least, despite the formal analogy with wormlike micelles, ($iii$) the physical origin of the characteristic timescale $\dot \gamma_c^{-1}$ remains an open issue.

From a theoretical perspective, the very existence of $\dot \gamma_c$ is still highly debated, even though it quickly emerged naturally from toy models based on the competition between shear and spontaneous physical aging \cite{Moller:2006,Coussot:2010}. The prevailing picture somewhat differs from the interpretation of non-monotonic flow curves in terms of a mechanical instability. It rather builds upon the generic observation that the static yield stress $\sigma_s$ above which the system starts flowing from the solid state is larger than the dynamic yield stress, i.e. the minimal stress at which the material is observed to flow when the shear rate is decreased from large values \cite{Picard:2002}. The flow curve can thus be seen as the superposition of a monotonic flow curve pointing towards the dynamic yield stress for vanishing shear rates, and of a static branch at $\dot \gamma=0$, for stresses below the static yield stress [\textbf{Figure~\ref{fig4}\textit{c}}]. This scenario, which does not necessarily imply the presence of an underlying decreasing flow curve below some critical $\dot \gamma_c$, is supported by simulations of Lennard-Jones glasses \cite{Varnik:2003,Tsamados:2010}, and by numerical work on concentrated soft particles \cite{Chaudhuri:2012b}. 

\begin{figure*}
\includegraphics[width=\linewidth]{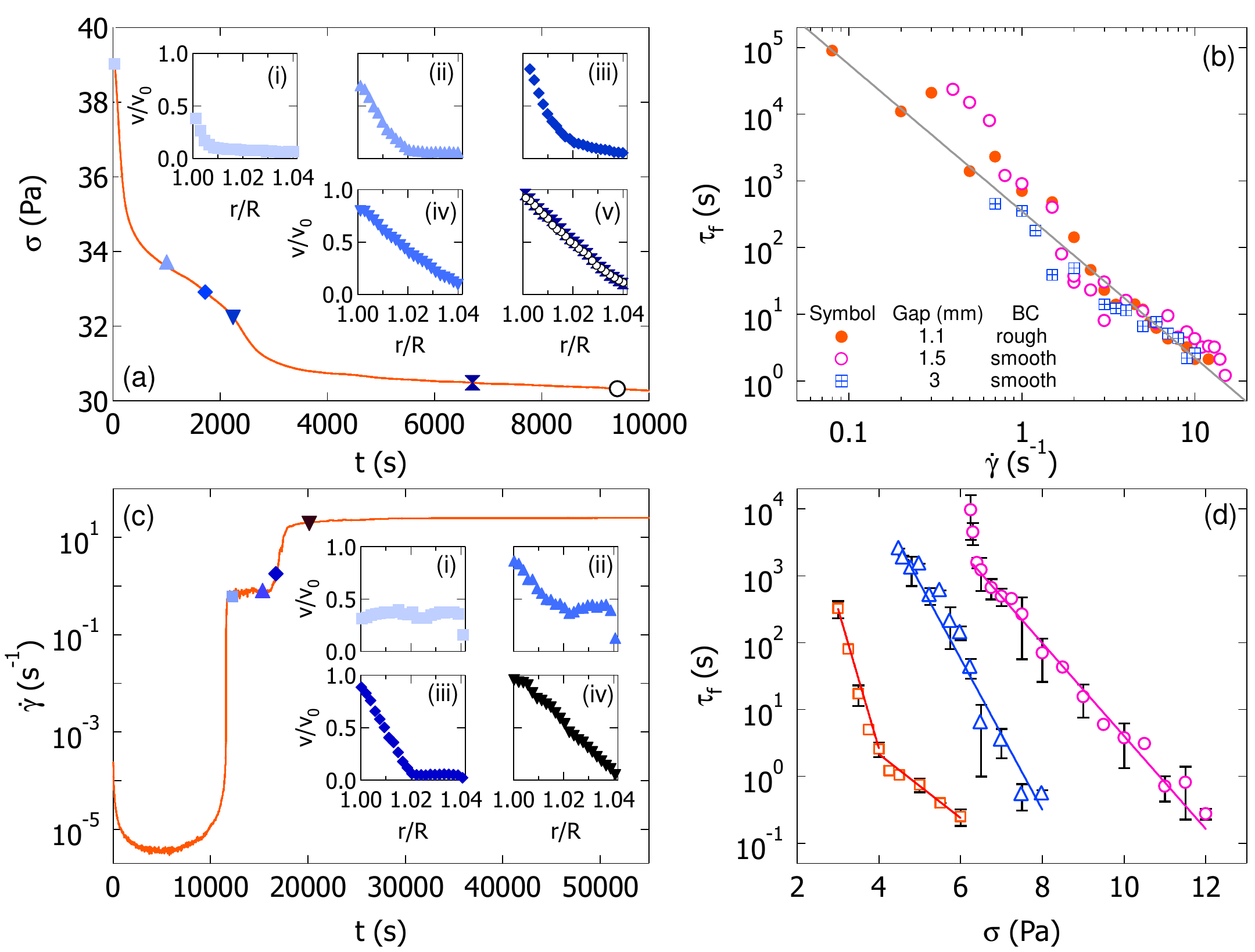}
\caption{(a) Stress response $\sigma(t)$ of a carbopol microgel to a shear start-up experiment at $\dot \gamma=0.7$~s$^{-1}$ in a small gap TC geometry with ${\rm R}=23.9$~mm. Insets: normalized velocity profiles $v/v_0$ at different times. The corresponding times are indicated in the main panel using the same symbols. (b) Fluidization time $\tau_f$ versus the applied shear rate $\dot \gamma$ for various gap widths and boundary conditions (BC). The gray line is the best power-law fit with an exponent $\alpha=2.3\pm 0.1$. Both (a) and (b) are adapted with permission from \cite{Divoux:2010}. Copyrighted by the American Physical Society. (c) Shear rate response $\dot \gamma (t)$ during a creep experiment in a 10\% wt. carbon black gel at $\sigma = 55$~Pa in a sand-blasted small gap TC cell of inner radius ${\rm R}=24$~mm. Insets: normalized velocity profiles $v/v_0$ at different times. The corresponding times are indicated in the main panel using the same symbols. Adapted from \cite{Grenard:2014} with permission of The Royal Society of Chemistry. Fluidization time $\tau_f$ vs the applied stress $\sigma$ for a thermoreversible gel of stearylated silica particles. The different curves correspond to gels that have been prepared by applying different levels of strain $\gamma_0$ prior to each creep experiment. From right to left: $\gamma_0=0$, 0.2 and 3.2. The lines correspond to exponential fits. Adapted with permission from \cite{Sprakel:2011}. Copyrighted by the American Physical Society.
}
\label{fig6}
\end{figure*}

In this framework the key role of attractive interactions has been mainly illustrated on emulsions made adhesive either by depletion forces \cite{Becu:2006} or by minute addition of clay particles bridging the droplets \cite{Ragouilliaux:2007,Paredes:2011}. However, the question of whether or not there is a minimal amount of attraction required to generate steady-state shear banding remains open. Likewise quantitative details on the potential link between the interaction potential and the critical shear rate $\gp_c$ are still lacking.

\subsection{From steady-state to transient heterogeneous flows}

Not all solid-like SGM display steady-state heterogeneous flows. Shear banding may be only transient and give way to a homogeneously sheared flow after long-lived induction periods. Such delayed, heterogeneous fluidization, now referred to as transient shear banding, has been reported for a large range of shear stresses above the yield stress on carbopol microgels \cite{Divoux:2010,Divoux:2011b}, carbon black gels \cite{Gibaud:2010,Grenard:2014} and laponite clay suspensions \cite{Gibaud:2008,Martin:2012}. The corresponding time-resolved scenario broadly involves several successive steps: homogeneous elastic deformation, strong (if not complete) slippage at the walls, nucleation and growth of a shear band that eventually spans the whole sample at a well-defined full fluidization time $\tau_f$ that dramatically increases in the vicinity of the yield stress \cite{Gibaud:2009,Divoux:2012} [\textbf{Figure~\ref{fig6}\textit{a,b}}]. In particular it has been established that $\tau_f$ diverges as power laws $\tau_f(\sigma)\propto 1/(\sigma-\sigma_c)^\beta$ and $\tau_f(\dot \gamma) \propto 1/\dot \gamma^\alpha$ in carbopol microgels \cite{Divoux:2010,Divoux:2011b} [\textbf{Figure~\ref{fig6}\textit{c}}]. Remarkably, in this system, which steady-state rheology follows the mostly phenomenological Herschel-Bulkley scaling $\sigma=\sigma_c+A\gp^n$ [\textbf{Figure~\ref{fig4}\textit{f}}], it could be shown that the exponent $n$ is linked to those of transient shear banding through $n=\alpha/\beta$. Such results strongly support interpretations in terms of critical phenomena and dynamical phase transitions \cite{Bocquet:2009,Divoux:2012,Chikkadi:2014}.

However, a general picture on transient shear banding is still out of reach and certainly depends on the microscopic details of the system. For instance, in spite of similar dynamics, $\tau_f$ in weakly attractive colloidal gels does not follow critical-like power laws but rather obeys exponential Arrhenius-like scalings [\textbf{Figure~\ref{fig6}\textit{d}}] hinting at a central role played by thermally activated processes \cite{Gopalakrishnan:2007,Gibaud:2010,Sprakel:2011,Grenard:2014}. Still, the microscopic ingredients at the origin of exponential {\it vs} power-law scalings involved in such delayed yielding remain to be uncovered, while predicting transient flow dynamics from microstructural information appears as an important open research topic \cite{Chaudhuri:2013}.

Despite these difficulties, a general criterion for transient banding has been recently formulated in the framework of trap and fluidity models by \cite{Fielding:2014}. Both approaches rely on the strong viscoelastic responses of YSF and predict the formation of transient heterogeneous flows during the stress relaxation that follows the stress overshoot observed after shear start-up \cite{Moorcroft:2013}. This approach is further supported by recent experiments on laponite suspensions \cite{Martin:2012}. 

\section{Conclusion}
In fluid dynamics and in continuum approaches in general, the formalism is set-up in such a way as to start from states that are usually assumed to be homogeneous and laminar, steady, and stable. Nevertheless, time and time again, experiments have shown that as soon as sufficient power is delivered into the material -for instance by applying shear- inhomogeneity, instabilities and transient or intermittent effects creep in the flow fields (velocity field, deformation rate field, stress field, concentration field, etc.). Viscoelasticity, yielding, flow instabilities and turbulence, flow-induced structures, shear banding and other types of shear localization, all these phenomena and more show us various aspects of a common fate. In this review, we have tried to isolate shear banding by laying out some of its salient properties across complex fluids. One such property is that it can be understood to a good extent by only referring to the mechanical fields of stress versus shear rate~: it can be understood as a ``material instability''. But as in any nonlinear context, identifying the cause from the effect is not always trivial. Non-monotonic flow curves and shear banding seem quite tightly connected, but they are not equivalent. Is shear banding causing the non-monotonic flow curve, or is the non-monotonic flow curve causing shear banding? In some cases the answer is not so clear cut, since wall slip, structural changes, concentration differences, or flow instabilities can also be at play.

In the case of polymeric fluids, we have seen how research on solutions of wormlike micelles provide a roadmap on how to disentangle some of the nonlinear effects from each other. Nevertheless, a lot of research remains ahead, and the controversial status of shear banding in polymers is a testimony of this state of affairs. Even the ``spurt'' effect that motivated the use of the term shear banding in fluids remains an open issue. Shear banding, wall slip and flow instabilities may all have some effects on this phenomenon.

As for soft glassy materials, we have seen that the presence of attractive interactions between the constituents, either controlled by van der Waals forces or produced by long-lived sticky contacts between the particles, leads to a rich and complex phenomenology where aging and time-dependence play a major role. Shear banding in attractive SGM intrinsically depends on the initial state and on the shear history as well as on the nature of the boundary conditions.

In both polymeric and soft glassy materials wall slip can no longer be described as a mere rheological artifact decoupled from bulk rheology as it used to be \cite{Barnes:1995,Buscall:2010}. For instance, recent space- and time-resolved studies have indeed shown that slippery walls foster fluidization against steady-state shear banding \cite{Gibaud:2009}. The chemical nature of the walls also modifies the flow profiles far away from the boundaries \cite{Seth:2012}. Such a strong influence of boundary conditions on both transient and steady-state flows is not yet captured by models and we believe that a major theoretical challenge for future years is to include the effects of boundaries at least through coarse-grained approaches if not at the microscopic level.

More generally the question of whether shear banding in polymeric systems and in SGM is of similar fundamental nature remains open. Strikingly, various models such as fluidity models seem to apply equally well to both kinds of systems and, up to minor adaptations, have been able to reproduce subtle effects of confinement on shear banding \cite{Masselon:2008,Goyon:2008} as well as transient regimes in polymeric and glassy systems \cite{Fielding:2014}. Although devising a general theory from a first-principle approach independently of the system details is probably not a realistic challenge, this suggests that the solid-fluid coexistence in SGM may be fundamentally not so far from the coexistence of flowing bands in polymeric fluids.

% Future Issues
\section*{FUTURE ISSUES}
\begin{enumerate}
\item Can shear banding in polymeric and soft glassy materials be understood within the same framework? The mechanical criterion proposed by Moorcroft and Fielding provides a promising approach, which remains to be tested fully in experiments.  
\item More microscopically, what are the relevant parameters (interaction potential, relaxation mechanisms, etc.) that generate the critical shear rates for shear banding ($\gp_1$, $\gp_2$ or $\gp_c$) and the value of the stress plateau?
\item In both polymeric and soft glassy materials wall slip seems quite tightly connected to shear banding. How could this be incorporated systematically in theoretical models? More generally, how do the chemistry and the deformability of the boundary conditions impact the fluid dynamics, especially during transient flows?   
\item What is the extent of the interplay between shear banding and elastic instabilities and/or concentration fluctuations? 
\end{enumerate}

%Disclosure
\section*{DISCLOSURE STATEMENT}
The authors are not aware of any affiliations, memberships, funding, or financial holdings that might be perceived as affecting the objectivity of this review.

% Acknowledgements
\section*{ACKNOWLEDGMENTS}
The authors acknowledge funding from the European Research Council under the European Union's Seventh Framework Program (FP7/2007-2013) / ERC grant agreement No. 258803, from Institut Universitaire de France and from the CNRS ``Theoretical Physics and Interfaces" PEPS project ``ComplexWall". We thank P. Coussot, S. Lindstr\"om and D. Vlassopoulos for kindly providing us with their original data.

\end{document}